\documentclass[namedreferences]{solarphysics}

\usepackage[hyperref,optionalrh,showbiblabels]{spr-sola-addons} 
\usepackage{graphicx}        
\usepackage{color}           
\usepackage{breakurl}        
\usepackage[dvipsnames]{xcolor}

\newcommand{\ie}{{\it i.e.}}
\newcommand{\eg}{{\it e.g.}}

\newcommand{\aap}{    {\it Astron. Astrophys.}}

\newcommand{\apj}{    {\it Astrophys. J.}}
\newcommand{\apjl}{   {\it Astrophys. J. Lett.}}

\newcommand{\jgr}{    {\it J. Geophys. Res.}}

\newcommand{\solphys}{{\it Solar Phys.}}
 
\newcommand{\ssr}{    {\it Space Sci. Rev.}} 
\newcommand{\asr}{    {\it Adv. Space Res.}} 
\chardef\us=`\_

\begin{document}

\begin{article}
\begin{opening}

\title{On flare-CME characteristics from Sun to Earth combining remote-sensing image data with in-situ measurements supported by modeling}

\author[addressref={aff1},corref,email={manuela.temmer@uni-graz.at}]{\inits{M.}\fnm{Manuela}~\lnm{Temmer}}
\author[addressref=aff1,email={julia.thalmann@uni-graz.at}]{\inits{J.K.}\fnm{Julia}~\lnm{Thalmann}}
\author[addressref=aff1,email={karin.dissauer@uni-graz.at}]{\inits{K.}\fnm{Karin}~\lnm{Dissauer}}
\author[addressref=aff1,email={astrid.veronig@uni-graz.at}]{\inits{A.M.}\fnm{Astrid}~\lnm{Veronig}}
\author[addressref=aff1,email={johannes.tschernitz@edu.uni-graz.at}]{\inits{J.}\fnm{Johannes}~\lnm{Tschernitz}}
\author[addressref=aff1,email={juergen.hinterreiter@edu.uni-graz.at}]{\inits{J.}\fnm{J\"urgen}~\lnm{Hinterreiter}}
\author[addressref=aff2,email={rodriguez@sidc.be}]{\inits{L.}\fnm{Luciano}~\lnm{Rodriguez}}

\address[id=aff1]{Institute of Physics, University of Graz, Austria}
\address[id=aff2]{Solar–Terrestrial Center of Excellence, SIDC, Royal Observatory of Belgium}

\runningauthor{Temmer et al.}
\runningtitle{Remote sensing and \textit{in-situ} CME characteristics}

\begin{abstract}
We analyze the well observed flare-CME event from October 1, 2011 (SOL2011-10-01T09:18) covering the complete chain of action -- from Sun to Earth -- for a better understanding of the dynamic evolution of the CME and its embedded magnetic field. We study in detail the solar surface and atmosphere associated with the flare-CME from SDO and ground-based instruments, and also track the CME signature off-limb from combined EUV and white-light data with STEREO. By applying 3D reconstruction techniques (GCS, total mass) to stereoscopic STEREO-SoHO coronagraph data, we track the temporal and spatial evolution of the CME in interplanetary space and derive its geometry and 3D-mass. We combine the GCS and Lundquist model results to derive the axial flux and helicity of the MC from \textit{in situ} measurements (Wind). This is compared to nonlinear force-free (NLFF) model results as well as to the reconnected magnetic flux derived from the flare ribbons (flare reconnection flux) and the magnetic flux encompassed by the associated dimming (dimming flux). We find that magnetic reconnection processes were already ongoing before the start of the impulsive flare phase, adding magnetic flux to the flux rope before its final eruption. The dimming flux increases by more than 25\% after the end of the flare, indicating that magnetic flux is still added to the flux rope after eruption. Hence, the derived flare reconnection flux is most probably a lower limit for estimating the magnetic flux within the flux rope. We find that the magnetic helicity and axial magnetic flux are reduced in interplanetary space by $\sim$50\% and 75\%, respectively, possibly indicating to an erosion process. A mass increase of 10\% for the CME is observed over the distance range from $\sim$4--20~R$_\odot$. The temporal evolution of the CME associated core dimming regions supports the scenario that fast outflows might supply additional mass to the rear part of the CME. 
\end{abstract}
\end{opening}

\section{Introduction}
     \label{S-Introduction} 

Since the launch of the Solar TErrestrial RElations Observatory \citep[STEREO][]{howard08}, the Sun-Earth distance range is well covered as never before. Having three eyes viewing the Sun from different vantage points, many new insights about the initiation and subsequent propagation of coronal mass ejections (CMEs) in interplanetary space could be gained. Novel methods on 3D reconstructions of CMEs \citep{thernisien09,mierla10}, and with that, more detailed studies with respect to CME associated solar surface phenomena (e.g., flares or large-scale waves) were pursued, that could largely improve the understanding of CMEs \citep[e.g.,][]{kienreich09,temmer10,patsourakos12,bein12}. \textit{In situ} measurements at 1~AU show signatures that can be related to CME-associated solar surface signatures as well as direct observations of CMEs in white-light. In this respect, simultaneous on-disk and off-limb observations provide an invaluable source of linking remote sensing and \textit{in situ} signatures. Due to their impact and potential geoeffectiveness, Earth-directed CMEs are of special interest. Using the unprecedented multi-viewpoint data sets currently available, we can enhance our knowledge on CME characteristics and their behavior in interplanetary space. Results obtained using on-disk imagery will provide valuable information for periods which are limited to single viewpoint observations.

The close relation between early CME evolution and relation to solar flares is well acknowledged \citep[e.g.,][]{zhang06,temmer08}. Often associated to flare-CME events are dark dimming regions observed as decreased emission in extreme-ultraviolet (EUV) and soft X-rays (SXR). These are most probably caused by the expansion and evacuation of plasma due to a CME, and are therefore interpreted as low-coronal footprints of CMEs \citep{hudson01}. The analysis of dimming regions is of special interest, as the plasma which is depleted from the corona may reflect the mass which is fed into the CME, maybe over hours \cite[e.g., see also][]{zarro99,harra01}. Hence, characteristic CME properties may be derived from the dimming evolution \cite[e.g.,][]{cheng2016}. \cite{qiu2007} derived the total magnetic reconnection flux in the low corona for flare-associated CMEs and their dimming regions, and compared it to the corresponding magnetic flux in magnetic clouds (MC) probed at 1~AU. For a sample of nine events they found that the reconnection flux in the flare is related to the magnetic flux of the MC. However, a straightforward comparison between flare characteristics, or dimming regions on the solar surface with off-limb measurements or \textit{in situ} counterparts is not an easy task as unknown processes causing the dimming \citep{mandrini2007} or projection effects from single spacecraft views \citep{dissauer16} may lead to erroneous interpretations.

Early studies linking filament and MC characteristics were successfully performed by \cite{bothmer98} who related interplanetary magnetic properties of MCs to filament orientation and handedness at the Sun. The helicity of an erupting flux rope is assumed to be conserved during the CME propagation in interplanetary space enabling us to link MCs observed \textit{in situ} to their solar sources \citep[e.g.][]{dasso05, rodriguez08}. With the power of multi-spacecraft data, revealing remote sensing as well as \textit{in situ} data from different vantage points, we are able to have an even more detailed look on the different aspects of CMEs, their interplanetary propagation behavior and associated \textit{in situ} signatures \cite[e.g.,][]{rodriguez11,kilpua13,moestl14}. More comprehensively, a variety of case-studies linked in more detail the different aspects of the Sun to Earth flare-CME events. E.g.\, \cite{moestl08} focused on the comparison between magnetic flux derived from flare reconnection and \textit{in situ} data. \cite{bisi10} performed an extensive study using multi-instrument data for the analysis of the CME-associated source region which was simulated from vector magnetic field data driven by artificial horizontal flux emergence. In a recent study, \cite{patsourakos16} tracked the cause of a strong space weather event, in particular focusing on the near-Sun magnetic field strength from which the geoeffectiveness might be assessed.

In the current study we investigate the centrally on-disk located CME-flare event from October 1, 2011 starting at 09:18~UT (SOL2011-10-01T09:18). Compared to already existing studies, we bring new aspects into the dynamic evolution of a CME and its embedded magnetic field, by analyzing in detail the solar source region using nonlinear force-free and finite-volume helicity modeling, deriving the reconnected flux from the CME associated flare ribbons and dimming areas. In a novel approach we attempt to combine model results from 3D reconstructions of the CME close to the Sun with \textit{in situ} models for obtaining the magnetic field characteristics of the associated MC. We compare the results derived from remote-sensing imagery and in-situ measurements and discuss the relationship between the parameters.

\section{Data and Methods}

We investigate in detail the flare-CME event from October 1, 2011. The CME event is launched from NOAA active region (AR) 11305 located at N10W08, associated with a M1.2 GOES class flare (start: 09:18~UT, minor peak 09:37~UT, major peak: 10:00~UT, end: 10:17~UT).  

\subsection{Flare energetics}\label{FlareE}

For the flare evolution, we study full disk H$\alpha$ filtergrams from Kanzelh\"ohe Observatory for Solar and Environmental Research (KSO) with a temporal cadence of about 6~sec covering the time range 09:18~UT until 11:00~UT \citep{poetzi15}. Together with the information of the magnetic field from the 720s LOS magnetogram of the \textit{Helioseismic and Magnetic Imager} aboard the \textit{Solar Dynamics Observatory} \citep[SDO/HMI][]{scherrer12,schou12,2014SoPh..289.3483H}, we derive magnetic reconnection rates from the separation of flare ribbons observed in H$\alpha$. The H$\alpha$ images are normalized and co-aligned to the first image (north-up and derotated to the reference time 09:18~UT). The magnetic field maps are binned to the pixel scale of the H$\alpha$ filtergrams using IDL (coreg\_map.pro). For the alignment between H$\alpha$ images and the magnetograms, HMI continuum images are used.

\begin{figure}
   	\centerline{\includegraphics[width=0.7\textwidth,clip=]{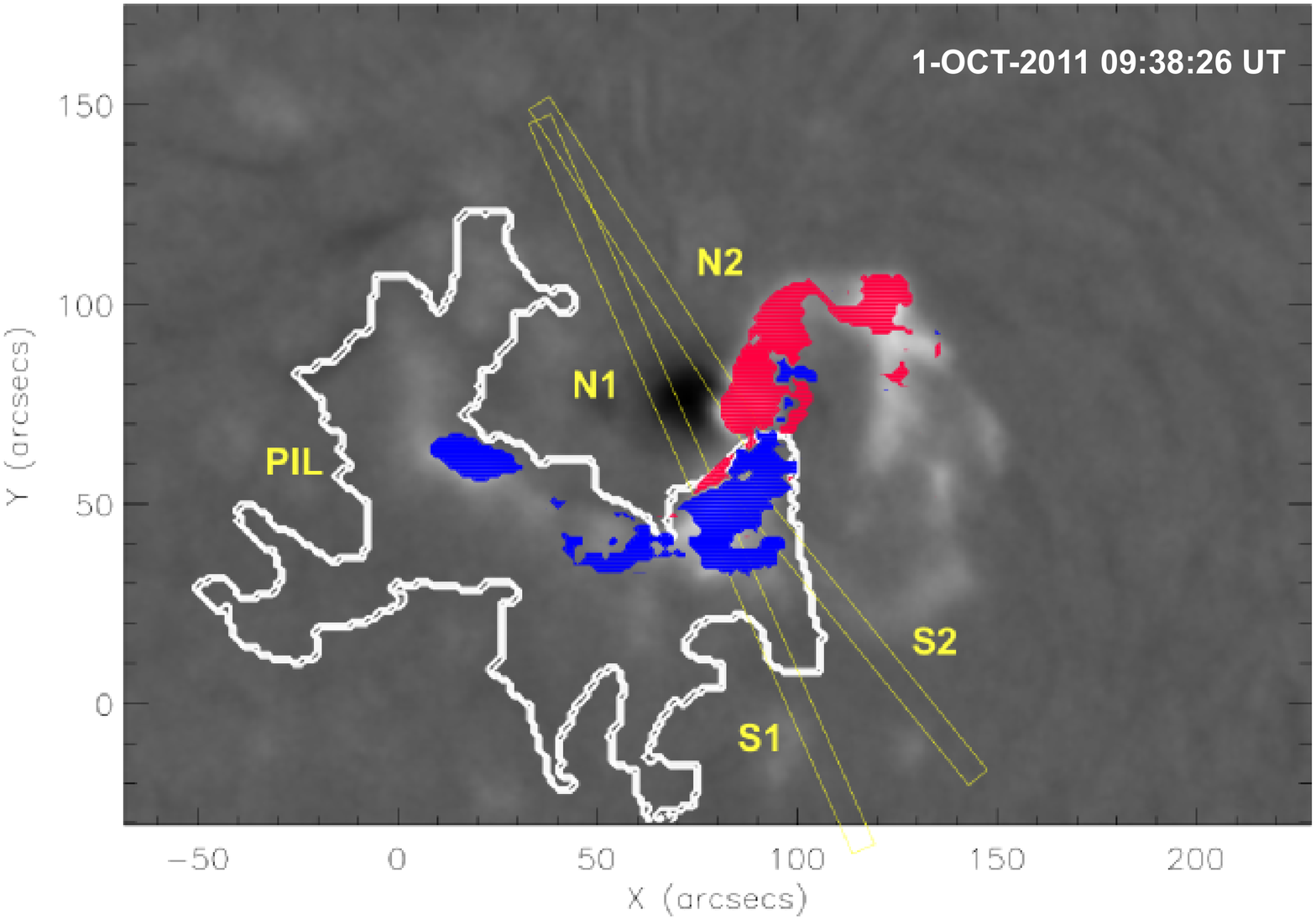}}
   	\caption{KSO H$\alpha$ filtergram showing the flare before it reaches its maximum intensity. The 
    polarity inversion line (PIL) is shown as white line, different directions (tracking paths: N1/2 for 
    the northern part and S1/2 for the southern part) along which the ribbon main motion is tracked are 
    shown with yellow rectangles, bright flare pixels cumulated until the time of the image shown 
    (09:38:26~UT) are shown as blue areas (positive polarity) and red areas (negative polarity).
    }
   	\label{ha}
   	\end{figure}

As shown in Fig.~\ref{ha}, we derive the flare ribbon separation speed from intensity profiles calculated along rectangular slices oriented perpendicularly to the PIL along two directions within each magnetic polarity (tracking paths: N1/2, S1/2). At each time step the intensity profile of each slice is fitted with a Gaussian function leading to a distance-time diagram. The time derivative of the polynomial fit of the derived distance-time curve is calculated to get the ribbon velocity and, hence, the local reconnection rate \citep{temmer07}.

The evolution of the flare ribbons provide us with important information on the coronal magnetic reconnection process in solar flare-CME events. Assuming translational symmetry in the flaring arcade that is built up behind the erupting CME, the reconnected electric field in the corona, $E_c$, can be derived from the local ribbon flare separation speed $v_r$ away from the polarity inversion line together with the underlying normal component $B_n$ of the photospheric magnetic field at the flare-ribbon location, as $E_c = v_r B_n$ \citep[cf.][]{priest86}. In case the flare does not occur too far off the disk center, the normal component $B_n$ can be well approximated by the LOS field as measured by SDO/HMI. \cite{ForbesLin2000} generalized this relation to three dimensions, showing that the rate at which magnetic flux is swept by the flare ribbons relates to a global reconnection rate. Assuming that the change of the photospheric field during the flare is small, this global reconnection rate can be determined from the observations as
\begin{equation}
\dot{\varphi }(t) = \frac{d\varphi}{dt} \approx \frac{\partial}{\partial t}\int B_n(a) da \, ,
\label{eq_magrec_2}
\end{equation}
with $da$ the newly brightened flare area at each instant and $B_n$ the normal component of
the photospheric magnetic field strength underlying $da$. This relation basically reflects the conservation of magnetic flux from the coronal reconnection site to the lower atmosphere, where the flare ribbons are observed \citep{ForbesLin2000}.

\subsection{Coronal dimming}\label{Sec:dimm}

We distinguish two different types of dimming regions, \textit{core} or \textit{twin} dimmings and \textit{secondary} or \textit{remote} dimmings \citep[see e.g.,][]{mandrini2007}. \textit{Core} dimmings are found in the form of stationary (long-lived) regions of strongly reduced EUV emission, and are closely located to the CME eruption site. Being located in regions of opposite magnetic polarity, they presumably resemble the cross-sectional area of the erupting flux rope footpoints at low-coronal heights. \textit{Remote} dimmings are observed over larger areas, extending to significant distances away from the eruption site, in the form of reduced EUV emission (though not as pronounced as in \textit{core} dimming regions).

We calculate the coronal dimming evolution from SDO/AIA \citep{pesnell12,lemen12} data in several wavelengths (most sensitive to quiet coronal temperatures around $\approx$ $0.6-2 \times 10^{6}$~K: 171~\AA, 193~\AA~and 211~\AA). The time series covers 12~h from the reference time 09:14~UT. We use high-cadence (12~sec) observations from 09:14 until 11:14~UT and a successively reduced cadence (1, 5, and 10~min) for the rest of the time series. The dimming regions are identified by applying a thresholding technique on logarithmically scaled base ratio images. A pixel is flagged as a dimming pixel if its logarithmic relative intensity is lower than $-$0.5 compared to its pre-event value. As an indication of the core dimming regions, we use the 10\% pixels in the dimming region that revealed the largest absolute change of their intensity below a certain threshold intensity. Naturally, applying the thresholding technique to the coronal images covering different temperature regimes (wavelength bands), results in different tracked extents of the dimming regions. Visual inspection suggests that the dimming areas are tracked best in 211~\AA\, which is also supported by previous studies \citep[e.g.,][]{robbrecht10,emil15}.

For the magnetic field information we use the 720s LOS magnetogram of SDO/HMI at the beginning of the event. All data were prepared using standard SolarSoft IDL software (aia\_prep.pro, hmi\_prep.pro), filtered for constant exposure time, and differentially de-rotated to the reference time. Based on this, we study the time evolution of the area of the coronal dimming regions and calculate the magnetic flux involved in the total dimming regions, only considering pixels with a magnetic field strength $|B_{i}|>$10~G, i.e.,\,above the HMI noise level. We note that values given for the dimming flux are derived from 211~\AA~image data using the arithmetic mean over positive and negative polarity.


\subsection{Coronal magnetic field modeling}

The 3D coronal magnetic field configuration in and around NOAA~11305 was modeled based on full-disk vector magnetic field observations from SDO/HMI. The ``hmi.B\_720s'' data series provides the total field, inclination and azimuth on the entire solar disk. The azimuth is provided with the 180$^\circ$-ambiguity already resolved in strong-field regions (using a minimum-energy method). For weak-field regions, we apply a so-called random disambiguation method, using the software tools provided by JSOC\footnote{The Joint Science Operations Center (JSOC) provides supporting documentation and software for HMI data. For details on the data series and related azimuth disambiguation procedure see \url{http://jsoc.stanford.edu/jsocwiki/FullDiskDisamb}.}. From the field, inclination and disambiguated azimuth, we retrieve the image-plane  components of the magnetic field vector, \ie, the LOS and transverse field. In order to account for projection effects, we de-project the image-plane data to a heliographic coordinate system, \ie, we derive the true vertical and horizontal field components, following \cite{1990SoPh..126...21G}. A sub-field of these optimized full-disk magnetic field data, covering the flaring AR as well as its nearest quiet-Sun surrounding, is used as an input to the nonlinear force-free (NLFF) coronal magnetic field modeling method \citep[for details see][and Sect.\ 2.2.1 of \citeauthor{2015ApJ...811..107D}~\citeyear{2015ApJ...811..107D}]{2010A&A...516A.107W}.\footnote{We list two important controlling parameters proposed in literature \citep[\eg,][]{2000ApJ...540.1150W,2006SoPh..235..161S}, in order to quantify the goodness of the obtained NLFF coronal magnetic field solution. For the current-weighted average of the sine of the angle between the modeled magnetic field and electric current density, we find $CWsin\approx0.1$. For the volume-averaged fractional flux we find $\langle|f_i|\rangle\approx10^{-4}$. (For a perfectly force-free and solenoidal solution, one would obtain $CWsin=0$ and $\langle|f_i|\rangle=0$.)} Using the 3D NLFF field as an input, we employ the finite-volume helicity method of \cite{thalmann11}, in order to estimate the relative helicity of the CME source region.

\subsection{CME morphology and kinematics}\label{CMEchar}

For deriving the entire kinematical profile of the CME evolution, we study combined EUV and white-light data from different vantage points using the SECCHI (\textit{Sun Earth Connection Coronal and Heliospheric Investigation}) instrument suite aboard the \textit{Solar TErrestrial RElations Observatory} \citep[STEREO;][]{howard08} as well as LASCO coronagraph data aboard SoHO \citep{brueckner95}. On October 1, 2011, the separation angle for STEREO-A--Earth and STEREO-B--Earth was 104.3$^\circ$ and 97.5$^\circ$, respectively, perfectly suited to derive reliable CME kinematics for an event centrally located on the solar disk from Earth view.

For estimating the CME geometry, its main propagation direction, and deprojected bulk speed, we use the graduated cylindrical shell (GCS) reconstruction method \citep{thernisien06,thernisien09}, which assumes an ideal flux rope to forward fit the appearance of the CME in coronagraph white-light data. The CME's early evolution is determined by manually tracking the frontal part of the CME along its main propagation direction, using SECCHI EUVI (with a field-of-view - FoV - up to 1.7~R$_\odot$), COR1 (FoV of 1.4--4.0~R$_\odot$) and COR2 (FoV of 2.5--15.0~R$_\odot$) image data from STEREO-A and -B. In addition we use the stereoscopic data in order to calculate the 3D-mass of the CME using the method described in \cite{colaninno09}. Following \cite{bein13} we derive the 3D CME mass evolution corrected for occulter effects over the distance range 1--20~R$_\odot$.

The interplanetary CME propagation from Sun to Earth orbit is tracked along the main propagation direction of the CME, applying the SATPLOT software tool\footnote{\url{http://hesperia.gsfc.nasa.gov/ssw/stereo/secchi/idl/jpl/satplot/SATPLOT_User_Guide.pdf}} available in IDL SolarSoft. The SATPLOT tool delivers j-maps for combined COR2, HI1 (FoV 4.0--24.0$^\circ$) and HI2 (FoV 18.7--88.7$^\circ$) white-light data making it easy to measure the elongation angle of the CME under study. The measured elongation-time profile is converted into a radial distance profile using the propagation direction and angular width, obtained from the GCS reconstruction. To obtain a range of possible propagation directions, we use several different conversion methods, including Fixed-Phi (FP), Harmonic Mean (HM) and Self-Similar Expansion (SSE), as described in \cite{sheeley99,lugaz09,davies12}, respectively. For calculating the CME speed and acceleration profile from the time-distance data, we apply the regularization method as described in \cite{temmer10}.

\subsection{\textit{In situ} CME characteristics}

To correctly identify the \textit{in situ} signatures of the CME (ICME) at Earth orbit, we perform drag-based-modeling (DBM) in order to simulate its interplanetary propagation along the main propagation direction \citep{vrsnak07zic,vrsnak13}. As an input, we use the CME's initial speed, distance, and angular width, as obtained from the GCS reconstruction. From the results we estimate the time range most suitable for studying the related ICME characteristics. We investigate the \textit{in situ} plasma and magnetic field by using 1-min resolution Wind data \citep{lin95,lepping95}. We apply a Lundquist force-free cylindrical fit \citep[hereafter referred to as Lundquist model; see][]{lundquist50} to the \textit{in situ} magnetic field data in order to reconstruct the properties of the ICME's flux rope, including its interplanetary orientation, radius and axial field strength. These are then used to calculate its axial magnetic flux and helicity, following \cite{devore00}.

\section{Results} 
\subsection{Source region characteristics: pre-flare structure}\label{ARprop}

	\begin{figure}[t]
	\centerline{\includegraphics[width=0.9\textwidth,clip=]{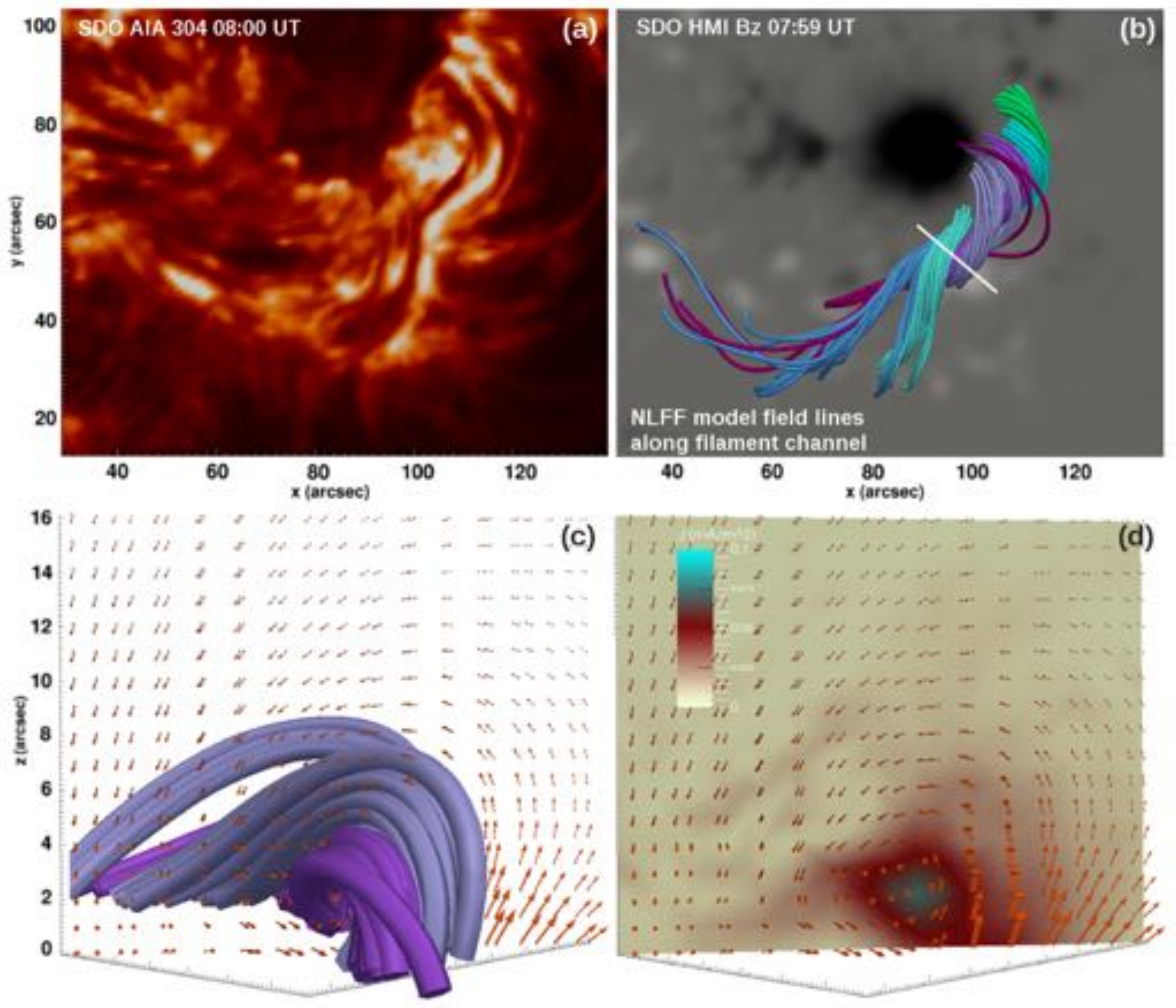}}
	\caption{(a) Central filament channel as observed in SDO/AIA 304~\AA, around the main sunspot of 
    NOAA~11305, prior to the flare-CME on October 1, 2011. (b) NLFF model magnetic field lines, 
    outlining the observed filament channel (colors are for better visibility only). The color coded 
    background resembles the SDO/HMI vertical magnetic field, scaled to $\pm 2~\mathrm{kG}$. (c) 
    Orientation of the coronal magnetic field (orange arrows) in a vertical cut through the model
    volume, above the path outlined as white solid line in (b). The light and dark violet model field
    lines are shown as in (b). (d) Orientation of the coronal magnetic field as in (c), but with the
    magnitude of the total electric current density shown as color-coded background.
    }
	\label{fig:frope}
	\end{figure}

The initial conditions for the eruption are derived from the pre-flare NLFF model (employed at 07:59~UT). The NLFF coronal magnetic model shows highly twisted magnetic fields, along the main PIL (Fig.~\ref{fig:frope}b), that clearly outline the dark filament observed in AIA 304~\AA\ (Fig.~\ref{fig:frope}a). Projected into a vertical plane, roughly perpendicular to the main axis of the filament, the coronal magnetic field vector exhibits a counter-clockwise pattern, \ie, a left-handed sense (orange arrows in Fig.~\ref{fig:frope}b,c). The NLFF model field lines (violet and purple lines in Fig.~\ref{fig:frope}c) warp around a central axis, at an approximate height of 3~arc-second ($\approx 2~\mathrm{Mm}$) above photospheric level, characterized by strongest electrical current density (Fig.~\ref{fig:frope}d). These properties are consistent with that of a coronal flux rope \citep[\eg,][]{2015JApA...36..157F}. The total unsigned axial flux within the flux rope is $1.1\times10^{21}$~Mx (estimated from the magnetic flux penetrating the vertical plane shown in Fig.~\ref{fig:frope}c and \ref{fig:frope}d). Using the 3D NLFF field as an input, we estimate the relative helicity of the AR core, hosting the flux rope, as $H_\mathcal{V}\approx -3.9\times 10^{42}~\mathrm{Mx}^2$. The relative helicity is a measure of how much a field is twisted and/or entangled, with respect to a reference potential field (of vanishing electric current and helicity). Its sign arises from the negative contribution of the left-handed fields to the AR's helicity budget \citep[\eg, review by][]{2007AdSpR..39.1674D}.


\subsection{Source region characteristics: eruptive phase} 
\subsubsection{Morphology of two-phase filament eruption}

   \begin{figure}    
   \centerline{\includegraphics[width=1.\textwidth,clip=]{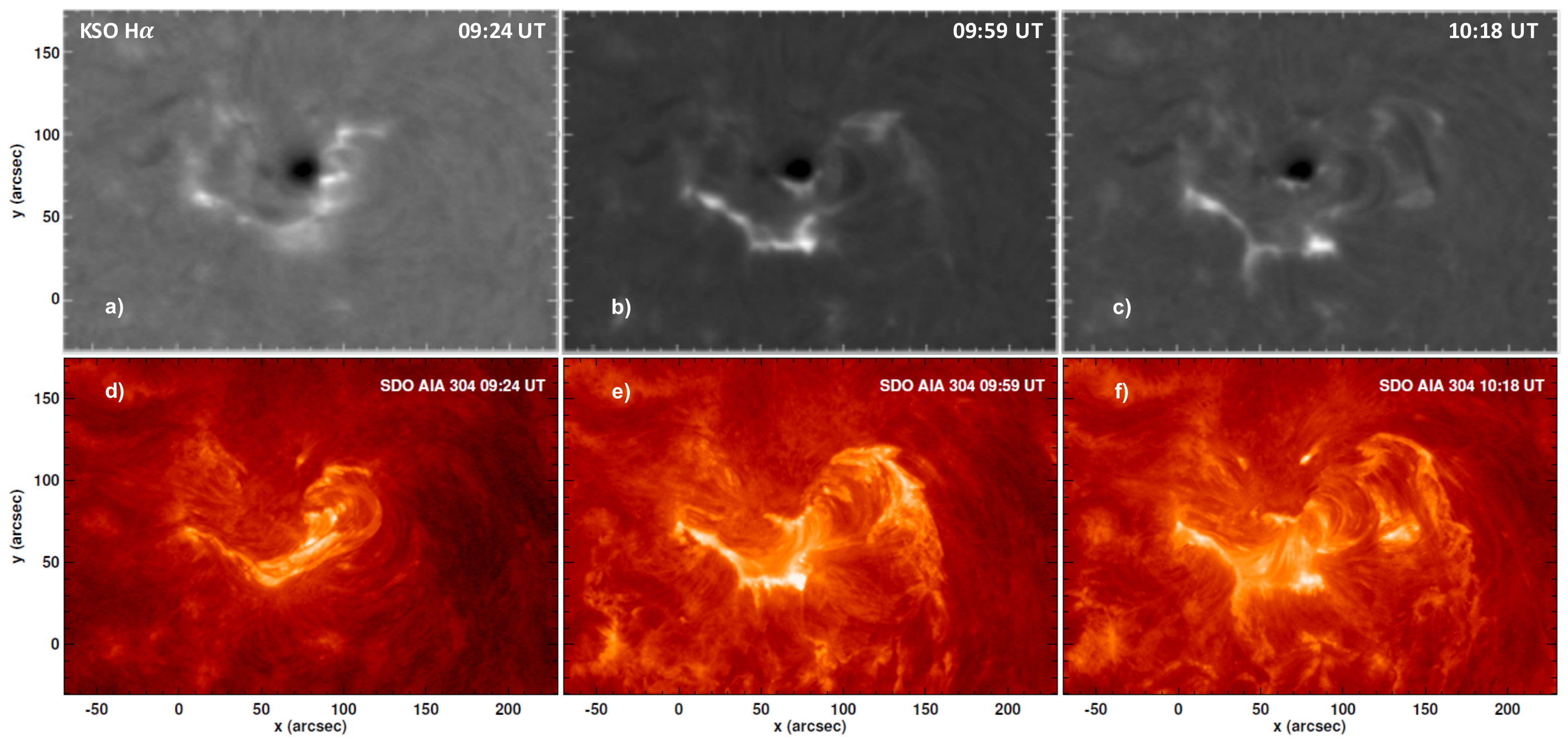}}
   \caption{Sequences of KSO H$\alpha$ (top panels) and AIA 304~\AA\ (bottom panels) images displaying the evolutionary stages of the flare, including the impulsive (left), peak (middle) and decay 
   (right) phase.
   }
   \label{ha-304}
   \end{figure}

Fig.~\ref{ha-304} shows the morphology of the flare as observed in H$\alpha$ (Fig.~\ref{ha-304}a--c) and AIA~304~\AA~(Fig.~\ref{ha-304}d--f). First, the filament to the south-west of the sunspot is activated and starts to rise around 09:22~UT (Fig.~\ref{ha-304}d), simultaneous to an initial rise of the observed SXR emission to C6 level (compare Fig.~\ref{ulti}g). In a second step, the structures in the south-east of the sunspot are destabilized (around 09:37~UT) and ending in the final eruption observed around $\sim$09:59~UT, co-temporal with the SXR emission rising towards the final M1.2-level. The flare ribbons observed in H$\alpha$ show a consistent evolution. Ribbon formation is observed first to the west of the sunspot (Fig.~\ref{ha-304}a), close to the location where the first post-flare loops become visible in EUV (compare \ref{ha-304}d), and evolves towards the south-east as the flare progresses (Fig.~\ref{ha-304}b).

\begin{figure}
	\centerline{\includegraphics[width=1.\textwidth,clip=]{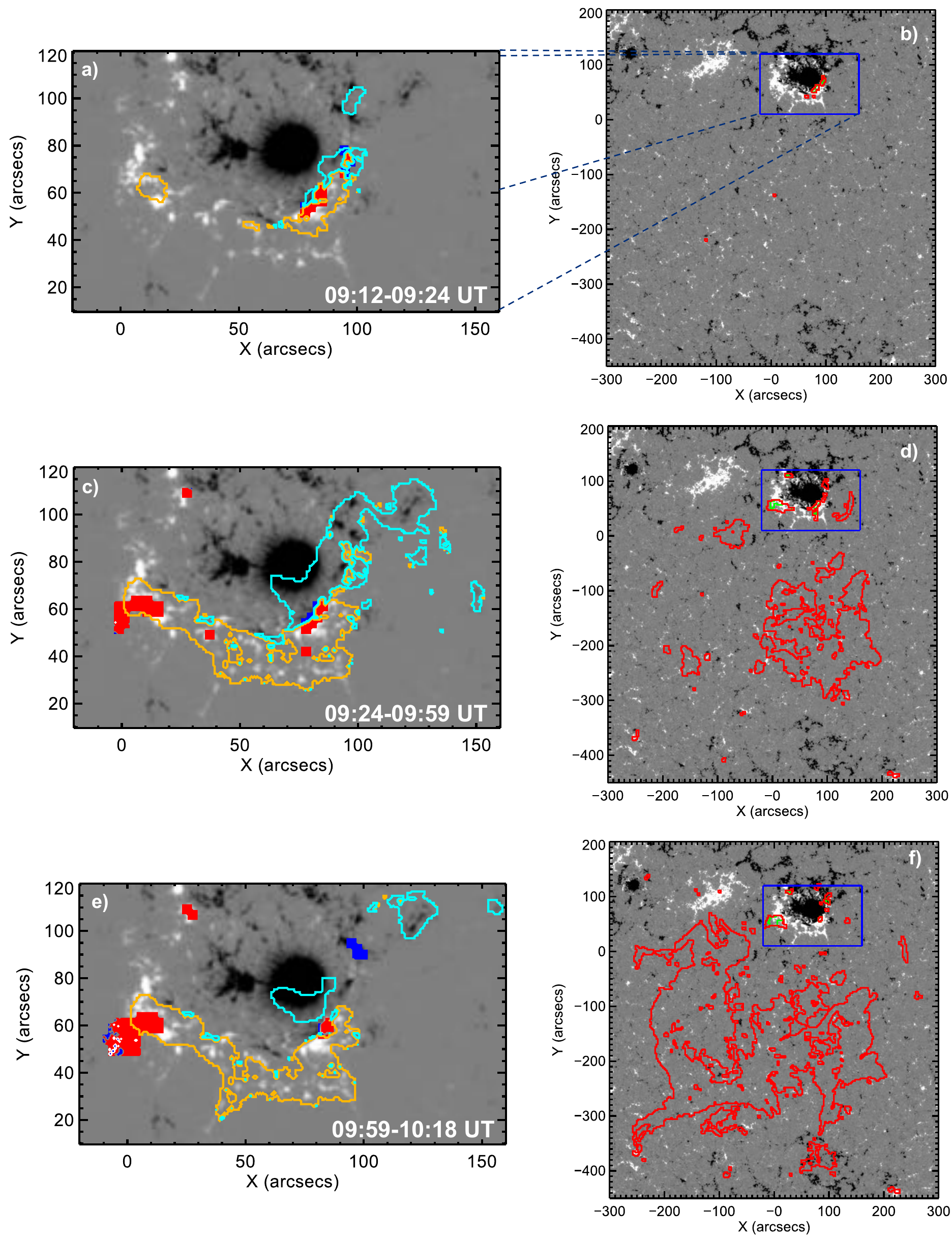}}
	\caption{Evolution of flare ribbons and coronal dimming, during three different time intervals covering the early impulsive phase (top), late impulsive phase (middle) and the decay phase (bottom) of the flare. \textit{Right panels:} Area covered by core (green filled
    contours) and remote (red contours) dimming. The blue rectangle outlines the flare region, shown in the left panels. \textit{Left panels:} Locations attributed to flare ribbons (cyan/yellow contours for
    signatures above negative/positive photospheric polarity) and 
    core dimming (blue/red filled contours above negative/positive polarity).
    The grayscale background resembles the HMI LOS magnetic field at 09:12~UT, scaled to $\pm$1 kG (left 
    panels) and to $\pm$0.1 kG (right panels) with black/white color representing the negative/positive
    polarity, respectively.}
	\label{fig:cdim_n_fpixels}
	\end{figure}

  \begin{figure}
  \centerline{\includegraphics[width=1.\textwidth,clip=]{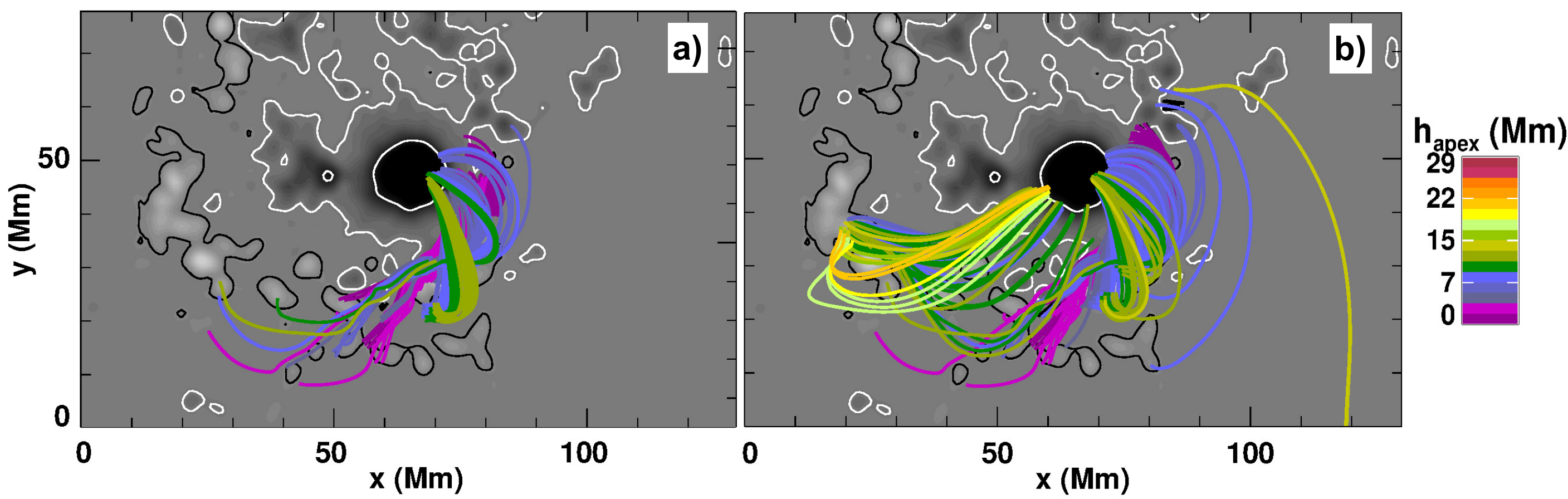}}
  \caption{(a) NLFF model field lines calculated from the core dimming regions, tracked between 09:12 and
  09:24~UT (see Fig.~\ref{fig:cdim_n_fpixels}). The field lines are color coded according their apex height. 
  The gray-scale background resembles the pre-flare vertical magnetic field, scaled to $\pm 2~\mathrm{kG}$. 
  (b) NLFF model field lines calculated from the flare pixels tracked in H$\alpha$ images for the same time
  interval as the core dimming shown in (b). Color-coding of the NLFF field lines and background as in (b). 
  Only field lines which close within the field-of-view and which connect photospheric regions of
  $B_z>10~\mathrm{G}$ are shown.}
  \label{fig:nlff_fl}
  \end{figure}
  
\subsubsection{Time evolution of flare-induced ribbons and CME-induced dimming}

\begin{figure}   
 	\centerline{\includegraphics[width=1.\textwidth,clip=]{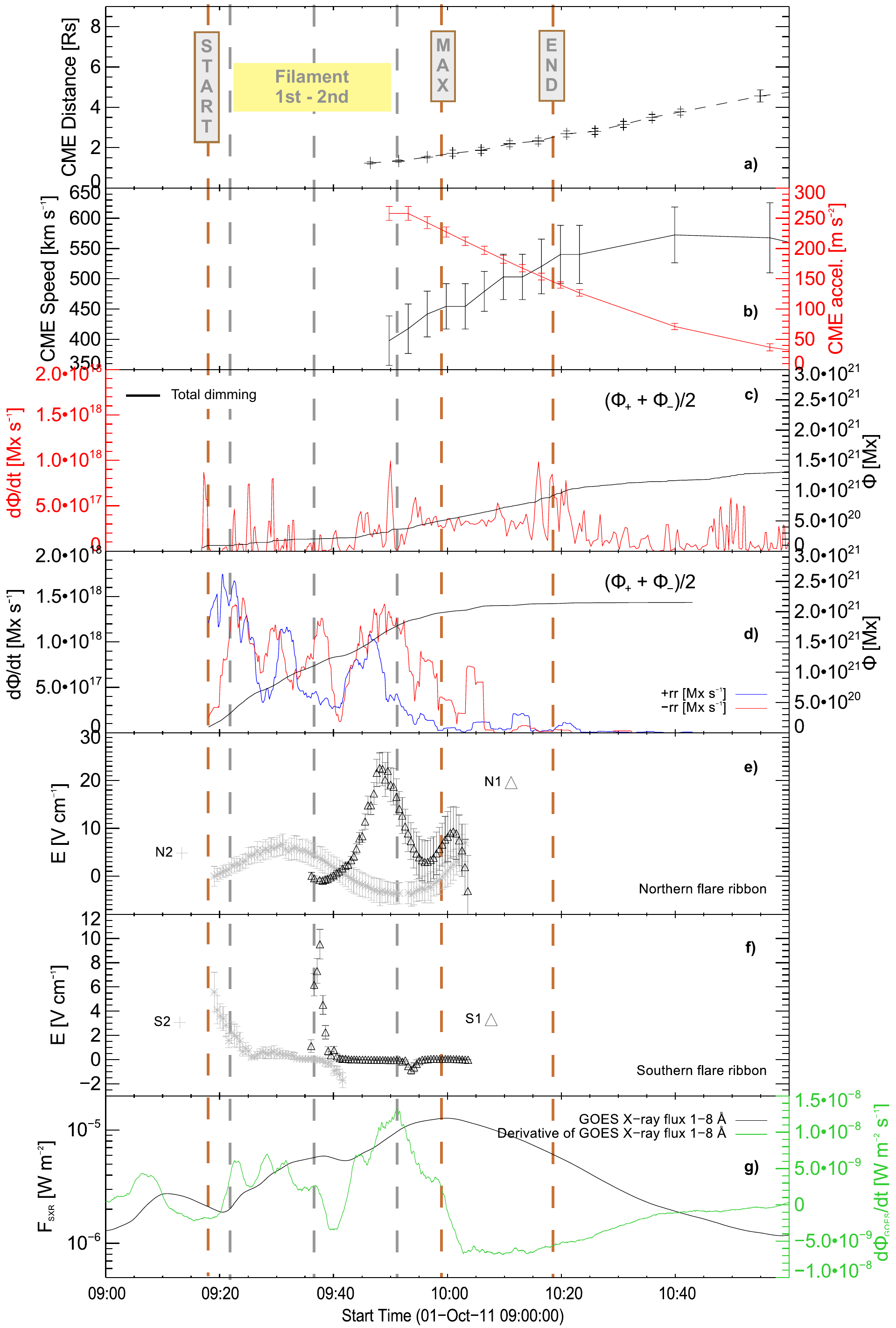}}
 	\caption{
    Time evolution of flare-/CME-related parameters, in comparison to the deduced CME kinematics.
    a) Distance-, b) velocity- and acceleration-time profile of the CME. c) Reconnected magnetic flux deduced from 
    dimming pixels (211~\AA) and d) from flare ribbon pixels covering areas of positive (blue) and negative (red) magnetic polarity. Local reconnection rate deduced along different directions of ribbon
    motion, normal to the local polarity inversion line, from the e) Northern and f) Southern flare 
    ribbon. g) GOES soft X-ray flux (black) and its time derivative (green). Yellow vertical dashed lines mark distinct times within the flare process, including the nominal flare start, peak (``MAX''), and end time. Gray vertical dashed lines mark different phases during the filament eruption. 
   	}
   	\label{ulti}
   	\end{figure}

In Fig.~\ref{fig:cdim_n_fpixels}, we trace both, the bright flare-induced ribbon emission and the diminished emission from the CME-induced dimming. From top to bottom, the signatures characteristic for the early impulsive (09:12 -- 09:24~UT), late impulsive (09:24 -- 09:59~UT), and decay (09:59 -- 10:18~UT) phase of the associated flare are shown. The panels of the right column outline the time evolution of the CME-associated coronal dimming, divided into core dimming and remote dimming (see Sect.~\ref{Sec:dimm}). In the panels of the left column, a close-up of the flare region (marked by the blue rectangle in the right column) is shown. Here, only the core dimming (blue and red filled contours) is shown, together with the locations populated by flare ribbon emission (cyan and yellow contours). During the early impulsive phase, flare ribbons and core dimming regions (Fig.~\ref{fig:cdim_n_fpixels}a) appear to the south-west of the sunspot, coinciding with the location of the filament observed before the flare (compare Fig.~\ref{fig:frope}a), and marking the footprint of the coronal magnetic field involved in the first phase of the flare. The flare ribbons and core dimming evolve towards the south-east of the sunspot only during the second phase of the flare (the late impulsive phase), coincident with the final eruption of the filament. With the launch of the CME, the  formation of pronounced and extended remote dimming areas is initiated (see Fig.~\ref{fig:cdim_n_fpixels}f and compare Fig.~\ref{ulti}a--c). As can be seen, this event reveals a complex interplay between flare brightened areas and core dimming regions. For this reason we use the core dimming areas only for qualitative purposes (cf.\,Sect.~\ref{mass:dimm}).

In Fig.~\ref{fig:nlff_fl}a we show NLFF model field lines traced from the detected core dimming area (cf.\, Section~\ref{Sec:dimm}) during the early impulsive phase (09:12 -- 09:24~UT). This allows us to infer some geometrical properties of the magnetic structure that later developed into the observed CME. It involves twisted fields (a flux rope; compare Fig.~\ref{fig:frope}) with apex heights of $\lesssim 7~\mathrm{Mm}$. A comparison with Fig.~\ref{fig:nlff_fl}b) shows the model field lines traced from the flare pixels tracked within the same time interval. Besides the low-lying magnetic flux rope to the south-west of the sunspot, also higher-reaching fields to its south-east (with apex heights up to $\approx 25~\mathrm{Mm}$) were subject to magnetic reconnection, demonstrating the magnetic connection between the different portions of the AR that were involved in the eruption. Therefore, we can assume that the flare-CME process, initiated in the form of a filament eruption to the west of the sunspot (coincident with the early impulsive phase), progressed to the south-east of the sunspot by destabilization of/reconnection with the overlying magnetic configuration in that part of the active region (marking the late impulsive phase of the flare). 

\subsubsection{Relative timing of flare- and CME-associated features}

Fig.~\ref{ulti} relates different flare- and CME-associated parameters to distinct phases during the flare. The two phases of the filament eruption, including the eruption of the filament to the west of the sunspot and the destabilization of the overlying field to the south-east, match well with the different phases observed in the GOES X-ray flux and its derivative (black and green curve in Fig.~\ref{ulti}g, respectively), the latter being a proxy for the flare-associated hard X-ray emission \citep{neupert68,veronig02}. The time derivative of the magnetic flux associated with the flare ribbon pixels (red and blue curve in Fig.~\ref{ulti}d, for positive- and negative-polarity associated pixels, respectively) reveals major changes throughout the impulsive phase. On the other hand, the time derivative of the magnetic flux associated with the dimming pixels (red curve in Fig.~\ref{ulti}c) suggests major changes for the time range covering the CME initiation until the CME attains maximum speed. At the end of the decay phase, when the SXR flux decreased again to C-level around 10:45~UT, we find a total accumulated reconnected flux from the flare ribbon evolution of 2.1$\cdot$10$^{21}$~Mx, approximately two times the flux involved in the dimming ($\approx$1.1$\cdot$10$^{21}$~Mx at 10:45~UT).

The local reconnection rate (Fig.~\ref{ulti}e and f; deduced from the flare ribbon separation velocity and associated magnetic flux; cf.\ Sect.~\ref{FlareE}) is distributed in a non-uniform way along the flare ribbon (compare the resulting curves along the two tracking paths N1 and N2 for the northern flare ribbon and along S1 and S2 for the southern flare ribbon, and see also Fig.~\ref{ha}) which has been observed also in earlier studies \citep[see e.g.,][]{temmer07}. The (velocity) acceleration time profile of the CME, as derived from combined EUV and COR1 measurements (cf.\ Section~\ref{CMEchar}), reveals a close relation with the time evolution of (the derivative of) the GOES X-ray flux. The reconnected flux associated with the flaring peaks first, followed by that associated with the dimming, followed by the CME's acceleration to its maximum speed ($\sim$550~km\,s$^{-1}$ at a distance of 4~R$_\odot$; see Fig.~\ref{ulti}b). 


\subsection{CME 3D characteristics and kinematical evolution}

\subsubsection{CME 3D-mass and near-Sun kinematics}\label{mass:dimm}

  	\begin{figure}
     \centerline{\includegraphics[width=1\textwidth,clip=]{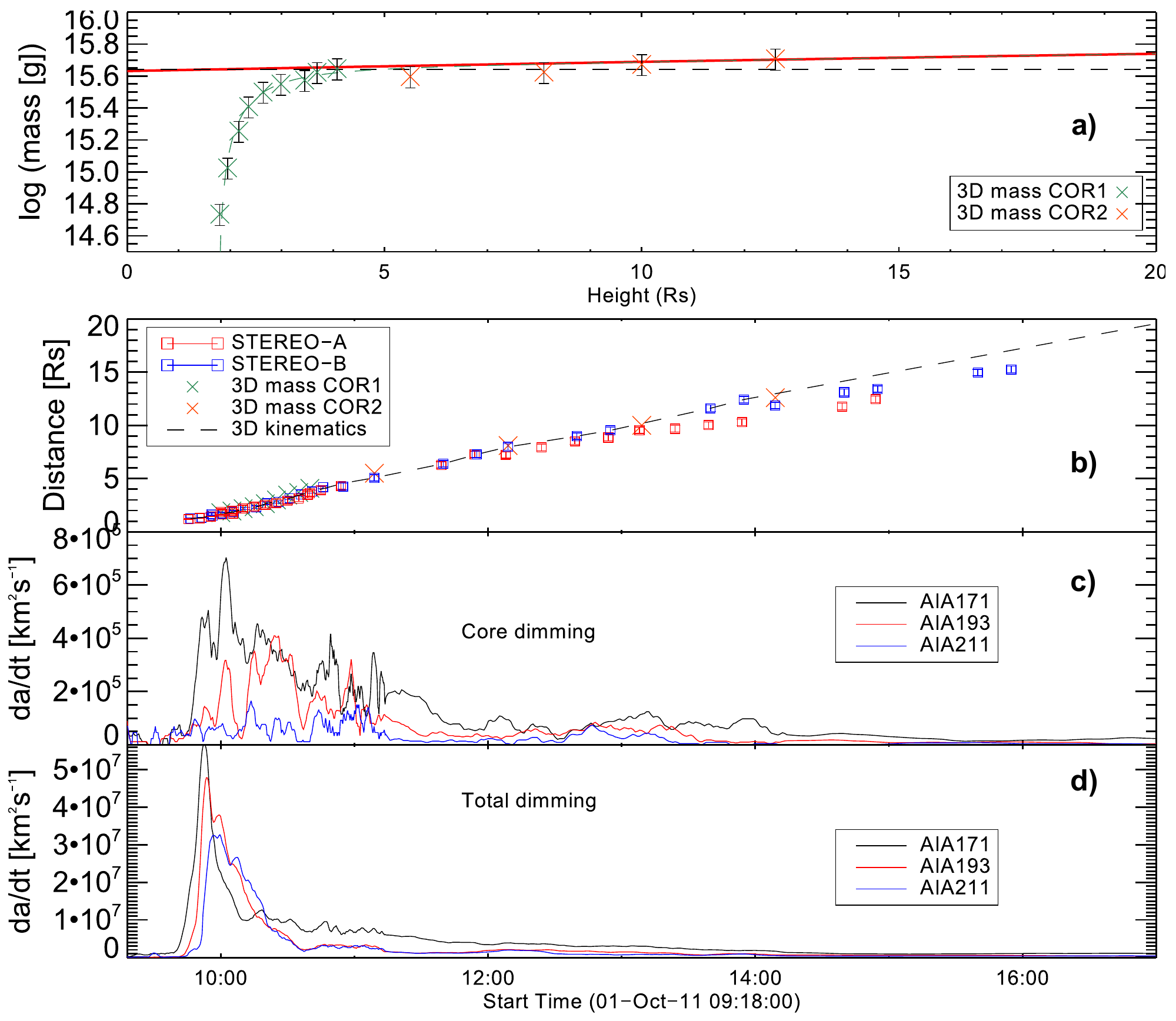}}
     \caption{\textit{Top to bottom panels}: Mass and kinematics of the CME within 20~R$_\odot$ above the solar surface. 
(a) 3D-mass estimate based on COR1 and COR2 observations. A fit (green dashed line) 
has been applied to the combined COR1 and COR2 measurements, where only those
masses estimated from COR2 have been used which exceed the COR1-based mass 
estimate at 4~R$_\odot$). Based on the fit, the 3D-mass evolution as function of height 
corrected for occulter effects (red solid line) and the seed-mass (horizontal black 
dashed line) is calculated. (b) CME distance-time evolution as derived from STEREO-A/B white-light images (red and blue squares, respectively) and the deprojected height of the CME front where 3D-mass measurements were made (green/red crosses). Co-temporal variation of the (c) core and (d) total dimming area measured from AIA 171, 193, and 211~\AA~image date. }  
     
     \label{mass}
     \end{figure}

Fig.~\ref{mass}a presents the near-Sun CME's 3D-mass evolution corrected for occulter effects. The CME kinematics up to a distance of $\sim$20~R$_\odot$ is given in Fig.~\ref{mass}b (cf.\,Fig.~\ref{ulti}a for the kinematical profile up to $\sim$5~R$_\odot$). The results suggest that the flare-associated ejection had a seed-mass of $m_0$=4.4$\cdot$10$^{15}$~g, that increased with a rate of $\Delta$m=6.1$\cdot$10$^{13}$~g~R$_\odot^{-1}$. As a result, we estimate the final mass at a distance of 20~R$_\odot$ as $m_{\rm end}\sim$5.5$\cdot$10$^{15}$~g.

Assuming that the observed increase in mass is due to a continuous mass flow that stems from the coronal regions where the CME footpoints are rooted, we compare the mass evolution to the evolution of the area covered by the CME-related coronal dimming (Fig.~\ref{mass}c,d). The time evolution of the total dimming area in three wavelengths (171, 193 and 211~\AA; see Fig.~\ref{mass}d), shows an effective growth starting at $\sim$09:45~UT that ceased around 10:20~UT. Relatively large variations in the core dimming area can be traced until $\sim$11:30~UT.


\subsubsection{CME geometry}\label{CMEShape}
    
    \begin{figure}
    \centerline{\includegraphics[width=0.9\textwidth,clip=]{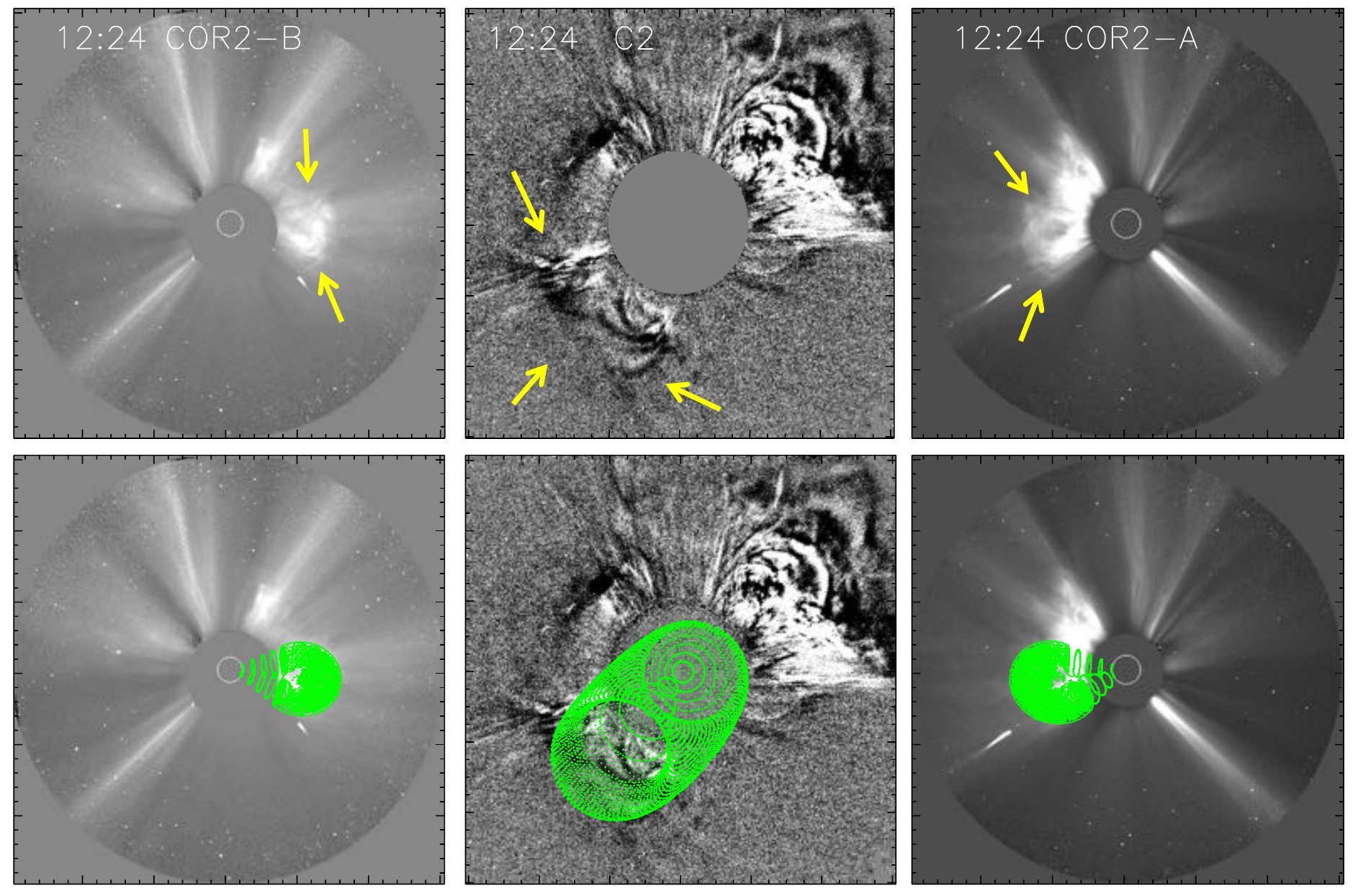}}
    \caption{GCS modeling results using the simultaneous view from three spacecraft (STEREO-B 	
    \textit{left}, 
    LASCO \textit{middle}, STEREO-A \textit{right}) on October 1, 2011. Yellow arrows mark the CME. The CME is 
    directed to south-east, with a clear tilt with respect to the ecliptic plane. The FoV 
    of LASCO C2 covers 2--6~R$_\odot$, STEREO-A/B 2.5--15~R$_\odot$.
    }
    \label{gcs}
  	\end{figure}

  \begin{table}[h]
  \caption{
  CME characteristics near the Sun and near Earth, using GCS modeling for the CME 
  geometry, 3D-mass calculation, and DBM to derive the CME's propagation characteristics in interplanetary
  space and arrival at Earth.
  }
  \label{table_gcs}
    \begin{tabular}{ll|ll}
     \hline
     \multicolumn{2}{c|}{Near Sun} & \multicolumn{2}{c}{Near Earth (1 AU)}   \\
     \hline
      GCS source region & E08S08 & GCS apex radius & 1.7$\cdot$10$^{12}$~cm (0.11~AU) \\
      GCS tilt vs.\ ecliptic & 45$^\circ$ & GCS flux rope length ($L$) & 3.9$\cdot$10$^{13}$~cm (2.6AU) \\
      GCS face-on width & 72$^\circ$ &  GCS volume & 9.0$\times$10$^{37}$~cm$^3$ \\
      GCS edge-on width & 34$^\circ$ & Density* & 25--35 cm$^{-3}$ \\
      GCS 3D speed & 450 km\,s$^{-1}$ & DBM arrival time & 2011-10-05 07:37~UT $\pm$ 5~h\\
      3D-mass  & 5.5$\cdot$10$^{15}$~g & DBM impact speed & 426~km\,s$^{-1}$ $\pm$ 30~km\,s$^{-1}$ \\
      \hline
    \end{tabular}
      *The density is calculated under the assumption that the mass stays constant beyond\\ 
      20~R$_\odot$ and is uniformly distributed within the derived CME volume. 
  \end{table}

Having three different vantage points, with almost perpendicular separation angles of the two STEREO satellites with respect to SoHO in the Sun-Earth line, we are able to reconstruct the 3D geometry of the CME from white-light coronagraphic data. The lower panels of Fig.~\ref{gcs} show the best fit (green cones), resulting from the GCS 3D flux rope model, when requiring that the boundary of the GCS model flux rope match the outer edge of the CME shape (indicated by yellow arrows in the upper panels) in STEREO-B (left), LASCO (middle) and STEREO-A (right) white-light images. At $t_0$=13:30~UT we obtain from the GCS model a CME distance of $r_0$=12~R$_\odot$, a propagation direction of $\phi_{\rm CME}$=$-$5$^\circ$, a speed of $v_0$=450~km~s$^{-1}$, and a CME half-width of $\lambda$=26$^\circ$. Note that due to the tilt of the reconstructed CME body ($\sim$45$^\circ$), we take an average of the face- and edge-on half-width. All parameters derived from the GCS modeling are summarized in Table~\ref{table_gcs}.

The obtained values are used as input for the DBM, in order to model the CME's interplanetary propagation. This allows us to compare distinct model parameters with actual \textit{in situ} measurements at Earth orbit. Using a drag value of $\gamma$=0.2$\times$10$^{-7}$ and an ambient solar wind speed of $w$=380~km\,s$^{-1}$, we estimate the ICME to arrive at Earth on 2011 October 5 at 07:37~UT ($\pm$ 5~h), with an impact speed of 426~km\,s$^{-1}$ ($\pm$ 30~km\,s$^{-1}$). Comparison with Wind observations allows us to determine the arrival of the CME-associated shock at 07:36~UT, with an impact speed of $\sim$460~km\,s$^{-1}$ and followed by a magnetic structure lasting from $\sim$10--22~UT (cf. Fig.~\ref{ulti_icme}). The ICME caused a moderate geomagnetic storm of \textit{Dst}=$-$43~nT \citep[``R\&C List''\footnote{\url{http://www.srl.caltech.edu/ACE/ASC/DATA/level3/icmetable2.htm}}]{richardson10}. The modeled and measured results show a quite good match revealing that the CME only marginally decelerated on its way from Sun to Earth. 

	\begin{figure}  
     \centerline{\includegraphics[width=1\textwidth,clip=]{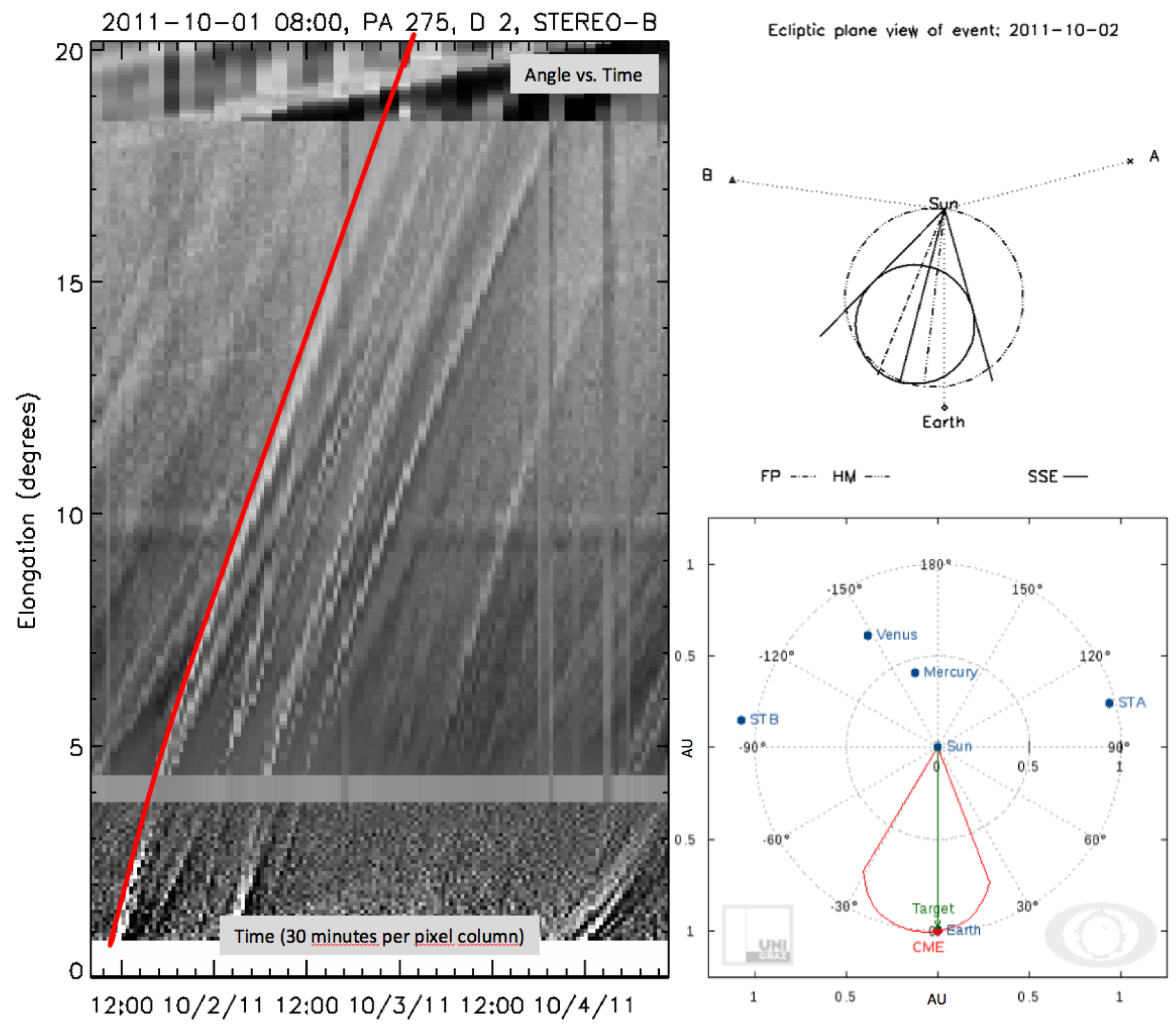}}
     \caption{\textit{Left}: Interplanetary propagation of the CME under study (red line) tracked using
     SATPLOT j-maps. \textit{Top right}: Conversion results from the derived elongation angle using
     several methods with different assumptions on the CME geometry (FP, HM, SSE - for more details see
     Sect.~\ref{CMEchar}). \textit{Bottom right}: DBM graphical output (\url{swe.uni-graz.at}) using as initial values the parameters derived from GCS model fit.}
     \label{satplot}
     \end{figure}
     
\subsubsection{Full kinematical profile}

     \begin{figure} 
     \centerline{\includegraphics[width=1.\textwidth,clip=]{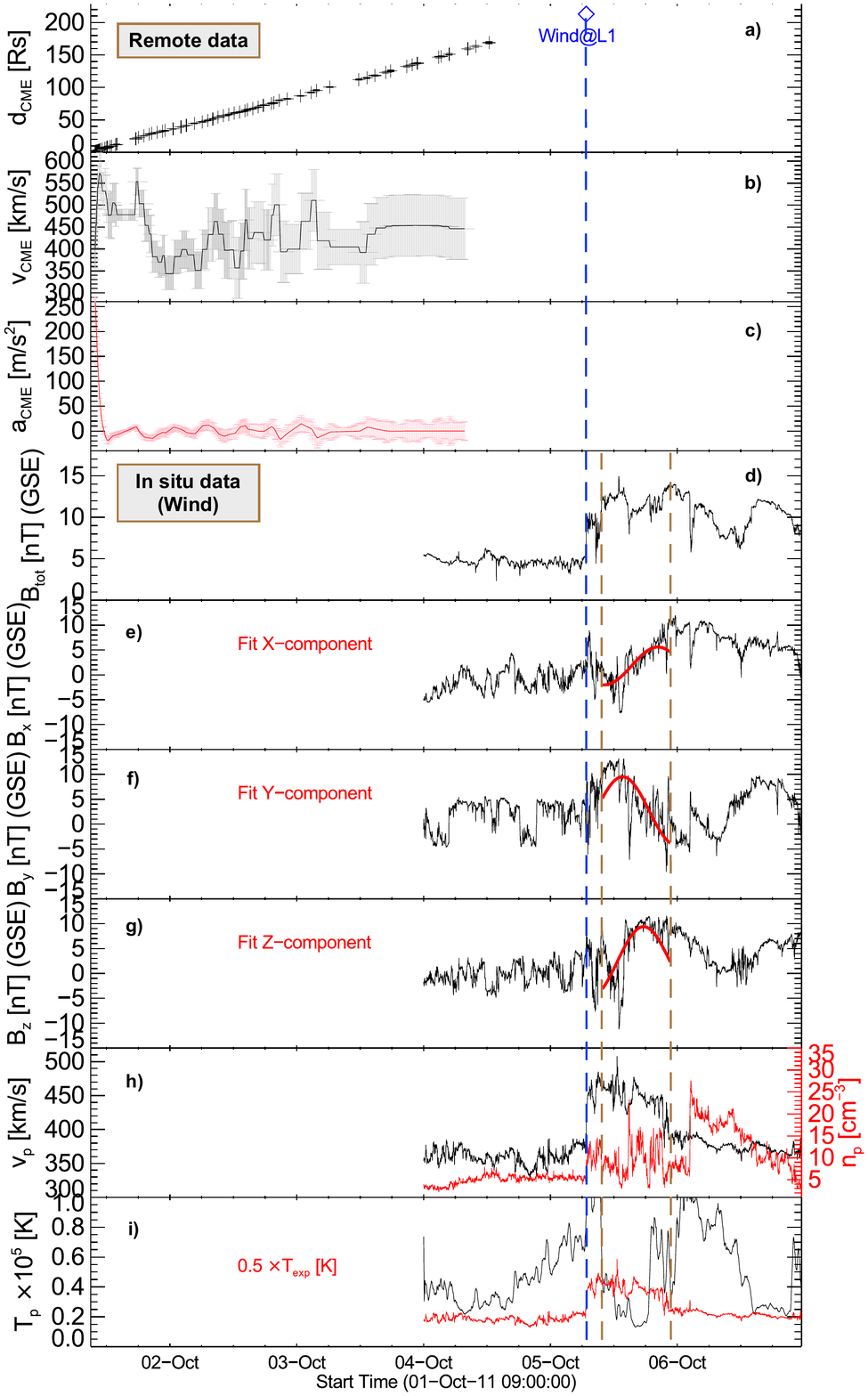}}
     \caption{\textit{Top three panels}: CME kinematics covering the distance range from Sun to the arrival of the CME at Wind spacecraft (blue dashed vertical line), including the (a) measured CME front height from the solar surface (black plus signs), (b) velocity, and (c) acceleration as a function of time. 
\textit{Bottom six panels}: In situ measurements
from Wind for the CME showing the magnitude (d) and direction (e--g) of the local interplanetary
magnetic field in GSE coordinates. The results of the Lundquist fit applied to the \textit{in-situ}
measurements of the MC covering the time span October 5, 2011 10:00--22:00 UT 
are indicated by red solid lines.
(h) Proton speed (solar wind bulk speed; black line), proton density
(red line), and (i) proton temperature (black line), together with the expected temperature
$T_{\rm exp}$ for quiet solar wind conditions (red line), based on which the extension of the
MC has been determined (brown dashed vertical lines).}
     \label{ulti_icme}
     \end{figure}

We were able to deduce the full kinematical profile of the CME all the way from the low solar corona up to 1~AU, based on combined EUV and white-light data. Fig.~\ref{satplot} shows the track of the CME in interplanetary space (covered by COR2, HI1, and HI2 data). By applying well-established fitting routines and assuming a constant propagation speed, we deduce the direction of propagation as E15$\pm$10 (SSE:$-$15$^\circ$, HM:$-$7$^\circ$, FP:$-$22$^\circ$; see top right panel in Fig.~\ref{satplot}). Importantly, this result is in accordance with the direction of propagation derived from GCS modeling, so that we can safely use the value of E15 for the conversion of the measured elongation angle to radial distances, and thus, for deriving the (I)CME kinematics, including the speed and acceleration profiles (see Fig.~\ref{ulti_icme}a--c and Section~\ref{CMEchar} for details). The CME front as observed in HI1+2 cannot be entirely tracked to the distance of L1, however, inspecting Fig.~\ref{ulti_icme}a, we see that a linear extrapolation of the derived kinematics would match well with the arrival of the CME at Wind spacecraft.


The plasma and magnetic field properties measured \textit{in-situ} by Wind, covering the time range 2011 October 4 00:00~UT to October 6 24:00~UT, are shown in Fig.~\ref{ulti_icme}d--i. They suggest that the CME shock-sheath structure arrived at Earth on October 5 at 07:36~UT (indicated by the blue dashed vertical line). Signatures typical for a MC \citep{burlaga91} were observed, including (i) a rotating magnetic field vector (between 10:00~UT and 22:00~UT; see Fig.~\ref{ulti_icme}e--g), (ii) an enhanced magnetic field strength (Fig.~\ref{ulti_icme}d), and (iii) a temperature below the typical quiet solar wind temperature \citep{richardson95}. Applying a Lundquist model to the \textit{in situ} measured data, we deduce an axial field strength of $B_{\rm 0}=$12.1~nT, a radius of the MC of $r_{\rm 0}=$1.75$\cdot$10$^{12}$~cm, and a relative orientation of the MC in interplanetary space (the axis of the embedded flux rope being inclined $\approx60^\circ$ with respect to the Sun-Earth line and $\approx54^\circ$ with respect to the ecliptic plane). Importantly, the inclination with respect to the ecliptic as well as the estimated radius of the MC agree with the corresponding value obtained from GCS modeling (cf.\,Table~\ref{table_gcs} and Section~\ref{CMEShape}). 

In an original approach, we combine the results obtained from the GCS modeling at 1~AU (cf.\,Table~\ref{table_gcs}) and the cloud parameters derived from Lundquist model of the \textit{in situ} data (cf.\,Table~\ref{table_insitu}), in order to compute the axial flux, $\Phi_{\rm ax}$, and helicity, $H$, of the MC following \cite{devore00}:

\begin{equation}
\Phi_{\rm ax}=1.4 \cdot B_0 \cdot r_0^2 ,\,
\end{equation}

and  

\begin{equation}
H=0.6 \cdot H_s \cdot B_0^2 \cdot r_0^3 \cdot L 
.\, 
\end{equation}

Here, $H_{\rm s}$ denotes the helicity sign, which is set to $-1$, corresponding to the left-handed flux rope deduced from the \textit{in situ} observed ICME signature (see Fig.~\ref{ulti_icme}d--i) and $L$ is the length of the MC that is calculated from the circumference of the GCS model result viewed face-on at 1~AU (see Table~\ref{table_gcs}). As a result, we obtain for the MC $\Phi_{\rm ax}$=5.2$\cdot$10$^{20}$ Mx and $H$=$-$1.8$\cdot$10$^{42}$ Mx$^2$, in basic agreement within a factor of two with the corresponding values derived for its source region on the Sun (summarized in Table~\ref{table_insitu} and see also Section~\ref{ARprop}). 

  \begin{table}[t]
  \caption{\textit{In situ} characteristics using a cylindrical force free model fit (see
  Fig.~\ref{ulti_icme}). For comparison we list the parameter values as derived from NLFF 
  model results and the solar source region studies.
  }
  \label{table_insitu}
  \begin{tabular}{lll}
    \hline
    \multicolumn{3}{c}{In situ measurements (Wind)}   \\
    Shock arrival time & 2011-10-05 07:36~UT & R\&C list    \\
    Impact speed & 460 km\,s$^{-1}$ &   R\&C list \\
    MC start & 2011-10-05 10:00 UT &   R\&C list \\
    MC end & 2011-10-05 22:00 UT &  R\&C list \\
    \hline
    \multicolumn{3}{c}{Lundquist model results*}   \\
    Axial field magnitude & 12.1 nT & $B_{\rm 0}$ \\
    MC radius (GSE) & 1.75$\cdot$10$^{12}$~cm (0.12~AU) & $r_{\rm 0}$  \\
    $\phi$ &  60.7$^\circ$ &  angle to Sun-Earth line\\  
    $\Theta$ &  54.1$^\circ$ & angle with ecliptic \\ 
    \hline
    \multicolumn{3}{c}{In situ CME magnetic characteristics}   \\
    $\Phi_{\rm ax}$ & 5.2$\cdot$10$^{20}$ Mx &  \\ 
    $H$ & $-$1.8$\cdot$10$^{42}$ Mx$^2$ &  \\  
    \hline
    \multicolumn{3}{c}{Solar flare-CME magnetic characteristics}   \\ 
    $\Phi_{\rm ax}$ & $1.1\times10^{21}$~Mx & NLFF \\ 
    $H_\mathcal{V}$ & $-3.9\times 10^{42}~\mathrm{Mx}^2$ & NLFF \\ 
    Ribbon flux (accumulated) & 2.1$\cdot$10$^{21}$~Mx &   H$\alpha$ (09:18--10:45~UT)\\ 
    Dimming flux (accumulated) & 1.1$\cdot$10$^{21}$~Mx  &   193~\AA (09:18--10:45~UT)\\
    Dimming flux (accumulated) & 1.4$\cdot$10$^{21}$~Mx &   193~\AA (09:18--11:30~UT)\\ 
    \hline
  \end{tabular}
  ~\\
  (*)We apply the Lundquist model to all \textit{in situ} data between 09:50 and 22:00 UT on\\ 
  October 05, 2011.
  \end{table}  

\section{Discussion and conclusion}

We study in detail the CME event from October 1, 2011. The analysis includes a wealth of data combining remote sensing and \textit{in situ} instruments to investigate the complete chain of action for the CME eruption and its evolution from Sun to Earth. We obtain detailed information on the solar surface signatures of the associated flare, magnetic field characteristics, and dimming regions that are subsequently  related to the \textit{in situ} plasma and magnetic field properties of the CME. 

The flare-CME event is associated with a filament eruption that actually occurred in two-steps, starting in the west of the source AR~11305 and moving towards south-east. The NLFF results can well explain the process and demonstrate the magnetic connection, showing that besides the low-lying magnetic flux rope to the south-west of the sunspot, also higher-reaching fields to its south-east were subject to magnetic reconnection. This is reflected in the CME propagation direction (E15) when compared to the source region coordinates (W08), and also in the location of the remote dimming regions. We derive that the magnetic flux rope of the CME is fed by two components, low-lying twisted magnetic fields (rooted in core dimming regions) and sheared overlying magnetic fields (rooted in flare pixels) involved in the eruption. We derive a flare reconnection flux of 2.1$\cdot$10$^{21}$~Mx and a dimming flux of 1.1$\cdot$10$^{21}$~Mx.

From H$\alpha$ emission we obtain the magnetic flux injected to the CME flux rope at different stages of eruption. We derive an equal amount of flare reconnection flux during the first impulsive phase of the flare (09:18--09:44~UT), i.e.\,before the SXR emission reaches M-level, and during the flare's major impulsive phase ($>$09:44~UT). Hence, reconnection processes were well ongoing before the filament started to erupt (09:37~UT) followed by the restructuring of the magnetic field. In comparison, the dimming flux, covering remote areas in the outskirts of the AR, shows regions involved in the reconnection process at a later time when the CME has already fully erupted (cf.\ bottom panel of Fig.~\ref{fig:cdim_n_fpixels}). Over the time range 10:45--11:30~UT, hence, after the flare has ceased, the dimming flux increased from 1.1$\cdot$10$^{21}$~Mx to 1.4$\cdot$10$^{21}$~Mx. This indicates that magnetic flux might have been added to the flux rope due to ongoing magnetic restructuring, too weak to produce visible H$\alpha$ flare ribbon emission. Therefore, the value of 2.1$\cdot$10$^{21}$~Mx for the total axial flux is most probably a lower limit. When comparing this to the \textit{in situ} axial magnetic flux of the MC ($\sim$0.5$\cdot$10$^{21}$~Mx) we find that it is reduced by at least 75\% and that the helicity is reduced by a factor of two. This might refer to an erosion of the MC while propagating in interplanetary space \citep[e.g.,][]{dasso06,ruffenach15}. For calculating the helicity from \textit{in situ} data, we took the best estimate of the MC length $L$, as derived from GCS modeling.

The determination of the magnetic flux \textit{in situ} as well as the MC radius $r$ and $L$ is prone to substantial errors. This is because the parameters used in its determination are derived from the fitting of an idealized magnetic field model (Lundquist force free cylindrical fit in our case) to \textit{in situ} data, where the selection of the MC boundaries affect the calculations. Furthermore, the \textit{in situ} models rely on a single 1D spacecraft crossing through a 3D structure, taking many assumptions into play \citep{demoulin16}. Also the GCS model is a fit of an idealized shaped CME to white-light data. In this respect we note that the MC radius, an important parameter to calculate the flux and helicity respectively, as determined from the \textit{in situ} model fit matches well with the radius as derived from the GCS extrapolated to 1~AU. For other case studies on this issue including poloidal flux components we refer to \cite{mandrini05,attrill06,qiu2007}.

The temporal profile of core dimming areas gives indication that the observed CME mass increase of 10\% in total, is supplied by the fast outflow from the core dimming regions. The CME mass consists of coronal plasma which gets compressed and moved away from the Sun due to the explosive release of magnetic energy. The associated dimming regions map the evacuation of the plasma, with the core dimming marking the CME footpoints and the remote dimming the CME body. In a qualitative approach we attempt to relate the temporal evolution of the core dimming and the CME mass increase. The plasma evacuated from the core dimming area would be detected in COR2 white-light only beyond the occulter radius of 2.5~R$_\odot$. Assuming an outflow speed of the order of 100--200~km\,s$^{-1}$ \cite[e.g.,][]{zarro99,harra01,tian12}, the plasma flow could be a detectable part of the CME mass after $\sim$1.5--3~h \citep[assuming a detection height beyond 4~R$_\odot$, see][this would yield 3--5.75~h]{bein13}. EUV observations reveal that the major changes in the core dimming ends around 11:30~UT, accordingly, outflows would feed mass into the CME until that time. Taking the propagation time as described above into account, this mass would become visible in the coronagraph white-light data latest around $\sim$17:15~UT. The CME apex is at a distance of about 17--18~R$_\odot$ at that time. This is consistent with statistical results showing that the increase in CME mass is primarily supplied to the rear-part of the CME to distances below 20~R$_\odot$ \citep[see][]{bein13}. This is supported by \cite{bemporad10} who observed at a distance of 4.1~R$_\odot$ continuous outflows in UVCS data over hours after the CME shock propagated through. They concluded that the transit of the CME flank left the coronal magnetic field open over $\sim$6~hours, facilitating fast plasma outflow before the corona recovered to the pre-CME configuration slowing down the outflowing plasma.

We calculate the CME density, using the GCS volume derived for the CME apex to be at 1~AU, and the observed 3D-mass at 20~R$_\odot$, which is assumed to be conserved. The derived density (25--35~cm$^{-3}$) is comparable within a factor of two to the \textit{in situ} measurements (10--15~cm$^{-3}$). However, the unknown plasma distribution in a CME volume still leaves many questions open, such as compression during the eruption and subsequent expansion as well as mass supply from fast outflows like core dimming regions. We note that the CME mass increase might continue during propagation in interplanetary space due to material swept up from the solar wind \citep[``snowplow effect'';][]{cargill04}. In the literature one finds quite high factors of the order of 2--3 for the CME mass increase in interplanetary space \citep[e.g.,][]{lugaz05,deforest13}. This might have effects on the drag that the CME experiences during its propagation phase. According to the DBM results from the event under study, the drag was of ``normal'' type and from the density estimate and comparison to the \textit{in situ} plasma density data, we could not confirm a substantial mass increase.  

Combining model data at various distance ranges gives us new insight into the CME characteristics as it propagates from Sun to Earth. However, the uncertainties especially in the derived magnetic field parameters and the lack of \textit{in situ} data at close distances to the Sun still leaves many questions open. New missions such as \textit{Solar Orbiter} or \textit{Solar Probe Plus} will deliver most eligible data sets to further pursue such studies. 

\begin{acks}
We thank the anonymous referee for helpful comments. The study was funded by the Austrian Space Applications Programme of the Austrian Research Promotion Agency FFG (ASAP-11 4900217) and the Austrian Science Fund (FWF): P24092-N16 and P25383-N27. J.K.T. acknowledges the excellent collaboration within the International team on Magnetic Helicity at the International Space Science Institute (ISSI, Bern). L.R. acknowledges support from the Belgian Federal Science Policy Office through the ESA-PRODEX program. The presented work has received funding from the European Union Seventh Framework Programme (FP7/2007–2013) under grant agreement No. 606692 [HELCATS]. This research was partially funded by the Interuniversity Attraction Poles Programme initiated by the Belgian Science Policy Office (IAP P7/08 CHARM). We thank A.~Gulisano and M.~Leitner for some helpful discussions. 
\end{acks}

\bibliographystyle{spr-mp-sola}

\end{article} 

\end{document}